\documentclass[%
 reprint,%
 amssymb, amsmath,%
 aip,cha,%
jap]{revtex4-1}

\usepackage{graphicx}
\usepackage{docs}%
\usepackage{bm}%
\usepackage[colorlinks=true,linkcolor=blue]{hyperref}%
\usepackage{textcomp}
\usepackage{placeins}
\expandafter\ifx\csname package@font\endcsname\relax\else
 \expandafter\expandafter
 \expandafter\usepackage
 \expandafter\expandafter
 \expandafter{\csname package@font\endcsname}%
\fi
\begin{document}

\title{Comment on ``Eshelby twist and correlation effects in diffraction from nanocrystals
`` [J. Appl. Phys. 117, 164304 (2015)]}%

\author{Jean-Marc Roussel} 
\email{jean-marc.roussel@univ-amu.fr}
\author{Marc Gailhanou}
\email{marc.gailhanou@univ-amu.fr}
\affiliation{Aix Marseille Universit\'e, CNRS, IM2NP UMR 7334, 13397, Marseille, France}
\date{\today}%

\begin{abstract}
The aim of this comment is mainly to show that anisotropic effects and image
fields should not be omitted as they are in the publication of A. Leonardi,
S. Ryu, N. M. Pugno, and P. Scardi (LRPS) [J. Appl. Phys. 117, 164304 (2015)]
on Palladium $\langle 011 \rangle$ cylindrical nanowires containing an axial
screw dislocation . Indeed, according to our previous study [Phys. Rev. B 88,
224101 (2013)], the axial displacement field $u_z$ along the nanowire exhibits
both a radial and an azimuthal dependence with a twofold symmetry due the
$\langle 011 \rangle$ orientation. $u_z$ is made of the superposition of three
anisotropic fields : the screw dislocation field in an infinite medium, the
warping displacement field caused by the so-called Eshelby twist and an
additional image field induced by the free surfaces. As a consequence by
ignoring both anisotropy and image fields, the deviatoric strain term used by
LRPS is not suitable to analyze the anisotropic strain fields that should be
observed in their Molecular Dynamics simulations. In this comment, we first
illustrate the importance of anisotropy in $\langle 011 \rangle$ Pd nanowire
by calculating the azimuthal dependence of the deviatoric strain term. Then
the expression of the anisotropic elastic field is recalled in term of strain
tensor components to show that image fields should be also considered.

The other aspect of this comment concerns the supposedly loss of correlation
along the nanorod caused by the twist. It is claimed for instance by LRPS that
: {\it``As an effect of the dislocation strain and twist, if the cylinder is
long enough, upper/lower regions tend to lose correlation, as if the rod were
made of different sub-domains.''}. This assertion that is repeatedly restated
along the manuscript appears to us misleading since for any twist the position
of all the atoms in the nanorod is perfectly defined and therefore prevents
any loss of correlation. To clarify this point, it should be specified that
this apparent loss of correlation can not be ascribed to the twisted state of
the nanowire but is rather due to a limitation of the X-ray powder diffraction
combined with the Whole Powder Pattern Modeling (WPPM). Considering for
instance coherent X-ray diffraction, we show an example of high twist where
the simulated diffractogram presents a clear signature of the perfect
correlation.
\end{abstract}

\maketitle


\section{Anisotropic strain field induced by an axial screw 
dislocation in a $\langle 011 \rangle$ fcc metal nanowire} The displacement
field $u_z$ induced by an axial screw dislocation in a $\langle 011 \rangle$
fcc metal nanowire has been studied in detail recently.\cite{MGJMR2013} For a
circular cross section, the $u_z$ field presents the two-fold symmetry of the
$\langle 011 \rangle$ orientation with an azimuthal $\theta$ dependence that
is controlled by the anisotropy of the shear modulus. This latter is
significant for Palladium since like in the case of Copper \cite{MGJMR2013}
the values of the elastic moduli are similar with $C_{44} \approx 28$ GPa and
$C_{55} \approx 82$ GPa in the $\{[100], [01\overline{1}], [011]\}$ coordinate
system.

\begin{figure}[ht!]
\begin{center}
\includegraphics[width=8.5cm]{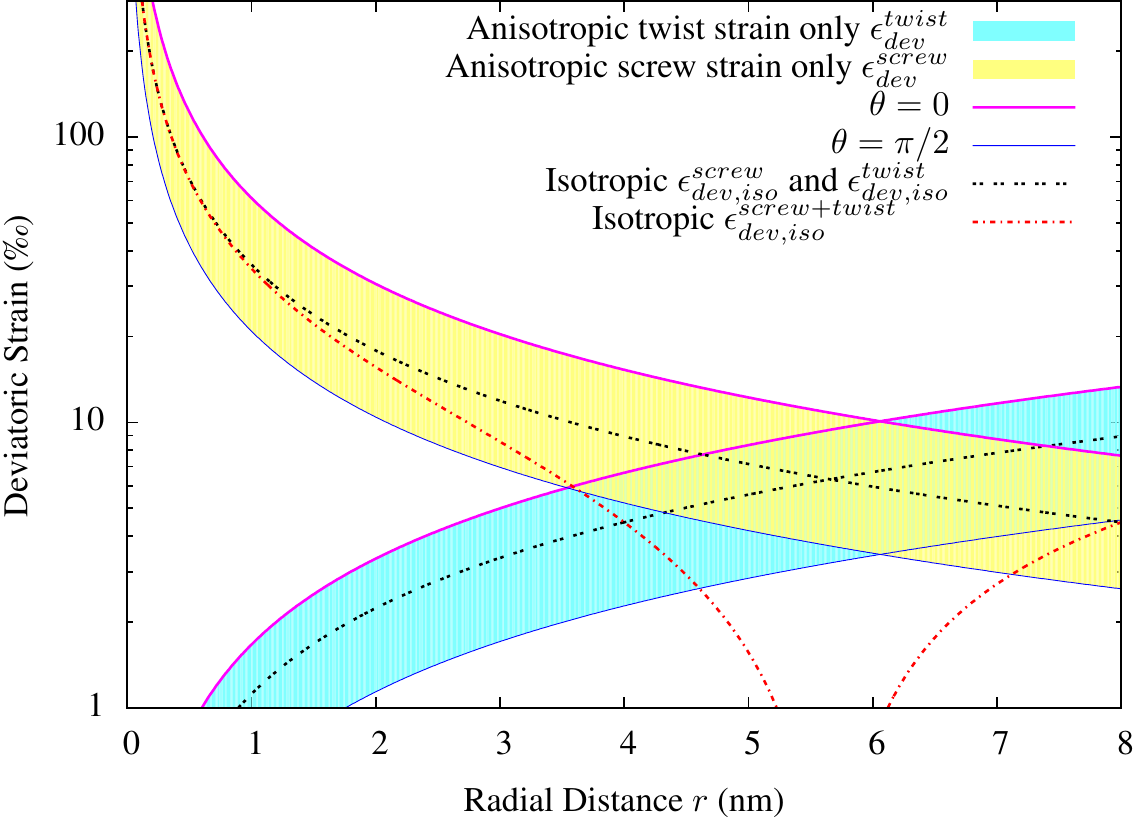}
\caption{Isotropic deviatoric strain terms reproduced from the Figure 6d of
LRPS\cite{LRPS} (black dotted lines) and compared to the same terms
$\epsilon^{screw}_{dev}$ and $\epsilon^{twist}_{dev}$ calculated from
anisotropic elasticity for all azimuth $\theta$ from
Ref.[\onlinecite{MGJMR2013}].  Clearly for Palladium, values of
$\epsilon^{screw}_{dev}$ and $\epsilon^{twist}_{dev}$ spread over large
domains bounded by extrema (for $\theta= 0$ and $\theta= \pi/2$) that differ
by a factor $C_{55}/C_{44}$.  The isotropic $\epsilon^{screw+twist}_{dev,iso}$
is also reported (red dotted line), it vanishes for $R/\sqrt{2}$.}
\label{devia}
\end{center}
\end{figure}

To illustrate the importance of these anisotropic effects, let us calculate
the azimuthal dependence of two particular quantities discussed in the article
of LRPS\cite{LRPS} (and reported in their Figure 6d), namely the deviatoric
strain term due to the screw deformation only and the one due to the twist
only. These latter, denoted here $\epsilon^{screw}_{dev}$ and
$\epsilon^{twist}_{dev}$ respectively, are plotted in Figure \ref{devia} as a
function of the radial distance $r$ but also for all azimuth $\theta$. 
Clearly the azimuthal exploration shows that both $\epsilon^{screw}_{dev}$ and
$\epsilon^{twist}_{dev}$ belong to large domains bounded by extremum values
(for $\theta = 0$ and $\theta = \pi/2$) that differ by a ratio equal to
$C_{55}/C_{44} \approx 2.93$.
\FloatBarrier

Incidentally, we wish to comment the analysis made of the deviatoric strain
terms in the Figure 6d, that leads the authors to conclude at the end of
section III: {\it ``..., so that the combined effect (screw and twist) gets
closer to the MD simulation.''}. This assertion is doubly misleading.  First
of all because the MD simulation curve must contain the above mentioned
anisotropy which is not shown on this graph (some clarification on the method
used to get the MD curve would be helpful). And secondly because the isotropic
deviatoric strain term designated as ``Screw and Twist deformation field'' in
Figure 6d does not match a calculation of the combined effect of both the
dislocation and the torsion. The plot of this term denoted as
$\epsilon^{screw+twist}_{dev,iso}$ in Figure \ref{devia} of the present work
reveals a very different behavior since $\epsilon^{screw+twist}_{dev,iso}$
should vanish for $r = R/\sqrt{2}$, $R$ being the nanowire radius. This result
can be directly understood by examining the $\epsilon_{\theta z}$ and
$\epsilon_{r z}$ strain components in this isotropic case: the $\epsilon_{r
z}$ are null for both the dislocation and the torsion but the
$\epsilon_{\theta z}$ components have opposite signs with
$\epsilon^{screw}_{\theta z,iso} = \frac{b}{4 \pi r}$ and
$\epsilon^{twist}_{\theta z,iso} = -\frac{1}{2} \frac{b r}{\pi R^2}$ where $b$
is the magnitude of the Burgers vector.\cite{1953:Eshelby,HuitreEtLotte}
Consequently, since $\epsilon_{dev,iso}
= \frac{4}{\sqrt{6}} \sqrt{\epsilon_{\theta z}^2 + \epsilon_{r z}^2}$, one
gets $\epsilon^{screw+twist}_{dev,iso}
= \frac{2b}{\pi \sqrt{6}} \sqrt{(\frac{1}{2r}-\frac{r}{R^2})^2}$. Thus, for
$r$ approaching $R/\sqrt{2}$ the combined effect (screw and twist) gets far
away from the MD simulations shown by the authors.

To conclude this section, we provide the expressions of the strain components
$\epsilon_{\theta z}$ and $\epsilon_{r z}$ leading to the anisotropic
behavior reported in Figure \ref{devia}.  We also derive from our previous
work \cite{MGJMR2013} the additional image strain field that results from the
interaction of the screw dislocation with the lateral surfaces of the
anisotropic cylinder.

Having determined the equilibrium stress components $\sigma_{\theta z}$ and
$\sigma_{r z}$ in Ref.[\onlinecite{MGJMR2013}], the derivation of the strain
field becomes straightforward by using the following relations:
\begin{align}
\epsilon_{\theta z} = & \frac{1}{2 C_{44} C_{55}}   
\bigg [ \  \  \  \ \sigma_{\theta z} c_{55}(\theta) - \sigma_{rz} c_{45}(\theta) \bigg ]\nonumber \\ 
\epsilon_{r z}     = & \frac{1}{2 C_{44} C_{55}}
\bigg [ - \sigma_{\theta z} c_{45}(\theta) + \sigma_{rz} c_{44}(\theta) \bigg ]
\label{derivuz}
\end{align}
where the elastic moduli can be written as $c_{44}(\theta) = C_{\oplus} +
C_{\ominus} \cos 2\theta$, $c_{55}(\theta) = C_{\oplus} - C_{\ominus} \cos
2\theta$ and $c_{45}(\theta) = C_{\ominus} \sin 2\theta$ with $C_{\oplus} =
(C_{44} + C_{55})/2$ and $C_{\ominus} = (C_{44} - C_{55})/2$.

Thus, the strain field induced by a perfect Volterra screw dislocation, with
Burgers vector b = 1/2 $a \langle 1 1 0 \rangle$ is inversely proportional to
$r$ with a marked $\theta$ dependence:
\begin{equation}
\epsilon^{screw}_{\theta z} =  \frac{b \sqrt{C_{44} C_{55}}}{4 \pi r c_{44}(\theta)}
\hskip 1.5cm  
\epsilon^{screw}_{r z}     =  0
\label{derscrew}
\end{equation}
The twist of the nanowire that is necessary to cancel the torque due to the
dislocation produces a $\sigma^{twist}_{\theta z}$ stress component
($\sigma^{twist}_{r z}$ is null for a circular cylinder) that in term of
strain becomes :
\begin{equation}
\epsilon^{twist}_{\theta z} =  \frac{-b r}{\pi R^2} 
\frac{c_{55}(\theta)}{C_{44}+C_{55}}
\hskip 0.5cm  
\epsilon^{twist}_{r z}     =  \frac{ b r}{\pi R^2} 
\frac{c_{45}(\theta)}{C_{44}+C_{55}}
\label{dertwist}
\end{equation}

Finally, in the present case of an anisotropic $\langle 011 \rangle$ nanowire
of circular cross section containing a coaxial screw dislocation, an image
stress field $\sigma^{img}$ is necessary to fulfill the condition of a
vanishing traction at the lateral surface. Formally, this condition reduces to
$\left. \sigma^{img}_{rz} \right|_{r=R}
+ \left. \sigma^{screw}_{rz} \right|_{r=R} = 0$ because $\sigma^{twist}_{rz}$
is null for a circular cross section.

Thus, looking for an image field that obeys both to the boundary conditions,
the equilibrium and the compatibility equations, we could obtain a numerical
solution of the stress field based on a Fourier series analysis.  Approximate
expressions of $\sigma^{img}_{\theta z}$ and $\sigma^{img}_{rz}$ were also
proposed in Ref.[\onlinecite{MGJMR2013}]. Using Eqs.(\ref{derivuz}), these
latter can be converted in term of strain and written as :
\begin{align}
  \epsilon^{img}_{\theta z} & = - \frac{b r}{4 \pi \sqrt{C_{44}C_{55}} R^2} \bigg [
c_{55}(\theta) \ln \bigg ( \frac{c_{44}(\theta)}{C_0} \bigg )
- \frac{c_{45}^2(\theta)}{c_{44}(\theta)} \bigg
] \nonumber \\ \epsilon^{img}_{rz} & = - \frac{b r}{4 \pi \sqrt{C_{44}C_{55}} R^2}
c_{45}(\theta) \bigg [ 1 - \ln \bigg ( \frac{c_{44}(\theta)}{C_0} \bigg
) \bigg ]
\label{imganalytic}
\end{align}
with $C_0$ is equal to $C_{55}/2$.
\begin{figure}[h!]
\hskip -0.5cm
\includegraphics[width=9cm]{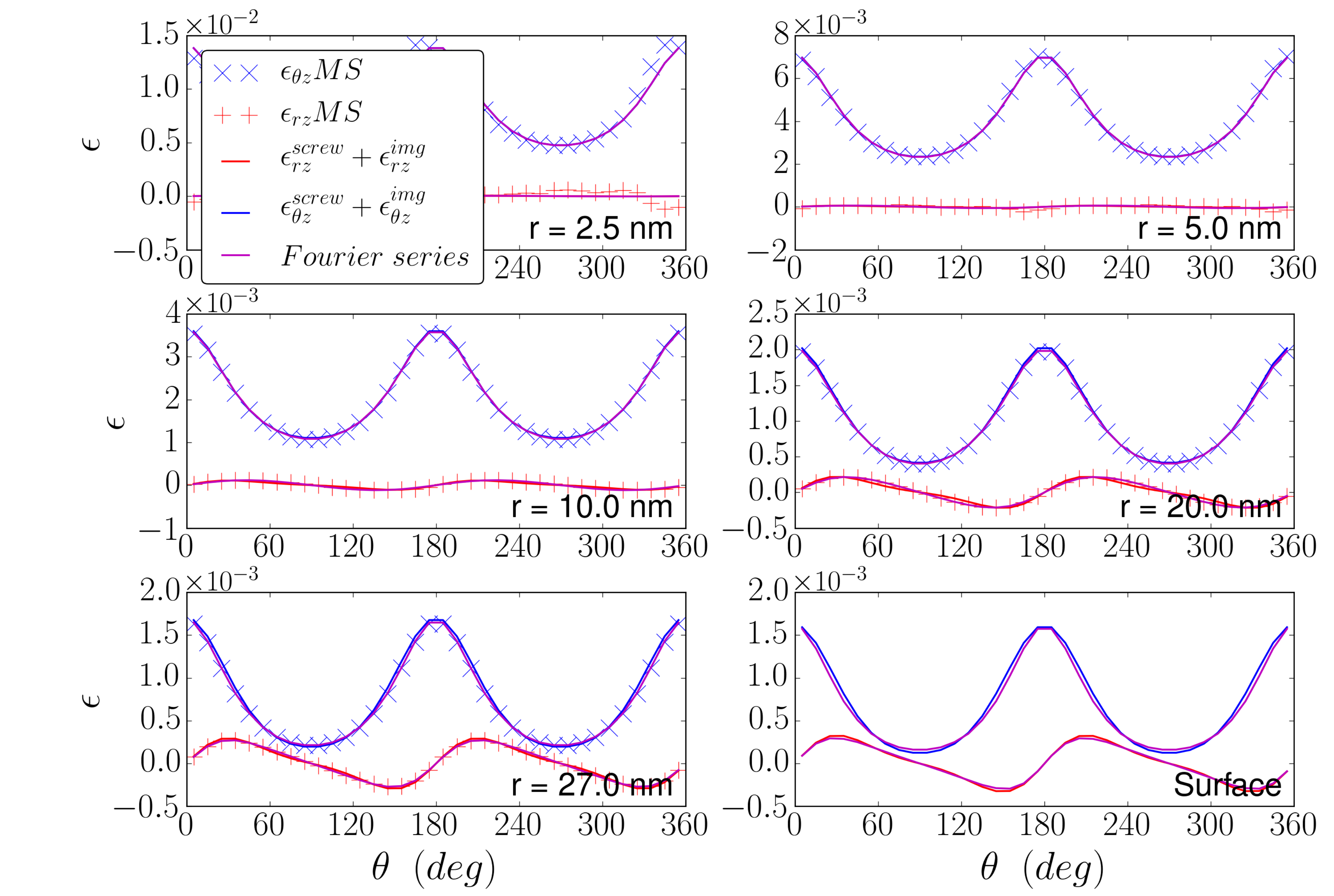}
\caption{$\epsilon_{\theta z}$ ($\times$) and $\epsilon_{rz}$ ($+$)
  strain components calculated from our Molecular Statics (MS) simulations in
  Ref.[\onlinecite{MGJMR2013}] at different $r$ values in the case of an
  untwisted [110] circular copper nanowire of radius $R = 30$nm containing an
  axial screw dislocation. These results are compared to the expressions of
  $\epsilon^{screw}_{\theta z} + \epsilon^{img}_{\theta z}$ and
  $\epsilon^{screw}_{rz} + \epsilon^{img}_{rz}$ (solid lines) proposed in
  Eqs. (\ref{derscrew}) and (\ref{imganalytic}). The boundary problem is also
  solved numerically through the Fourier series analysis described in
  Ref.[\onlinecite{MGJMR2013}].}
\label{deformation}
\end{figure}

In Ref.[\onlinecite{MGJMR2013}], the image field derived in term of stress
components was compared to the one calculated from Molecular Statics
simulations (MS). Similarly in the present comment, the MS simulations can
serve as a reference for testing the validity of the approximate expression
given in Eqs.(\ref{imganalytic}) of the image strain. In practice, the
analyticity of the Tight Binding potential used in our atomistic simulations
allows a straightforward determination of the strain components per atom. This
is illustrated in Figure \ref{deformation} where the radial and the azimuthal
dependencies of the strain field components $\epsilon_{\theta z}$ and
$\epsilon_{r z}$ resulting from our MS simulations are plotted in the case of
an untwisted Cu nanowire of radius 30 nm containing a screw dislocation at its
center (the torsion can be treated separately since it does not affect the
image field for a circular cross section). As for the stress analysis, the
same conclusions can be drawn. The dislocation field in Eq. (\ref{derscrew})
combined with the image field in Eq. (\ref{imganalytic}) capture well the
radial dependence and the azimuthal anisotropy of the strain field found in
our simulations.  This anisotropy is particularly pronounced for Copper (as
for Palladium). It controls for instance the shape of the Eshelby potential
well that traps the screw dislocation at the center of the twisted
nanowire.\cite{JMRMG2015}

\section{Diffraction from a twisted cylinder}

At the end of their article, LRPS arrive at the conclusion that {\it ``the
twist weakens the correlation between more distant regions of the cylindrical
domain, up to the point that needle-like nanocrystals appear as made of
sub-domains (...) which scatter incoherently.''}.

Fundamentally, torsion does not introduce any randomness of the atomic
positions and therefore can not be the cause of a loss of correlation. 

We believe rather that this apparent loss of correlation should be presented
as a limitation of the technique employed (i.e., the WPPM analysis combined
with X-ray powder diffraction) that does not permit to discern if the above
sub-domains scatter coherently or incoherently in such twisted samples.
Besides, it is worth completing that there are other techniques like X-ray
coherent diffraction that are capable to show the interference phenomena that
occur from the different sub-domains.

\begin{figure}[ht!]
\includegraphics[width=7cm]{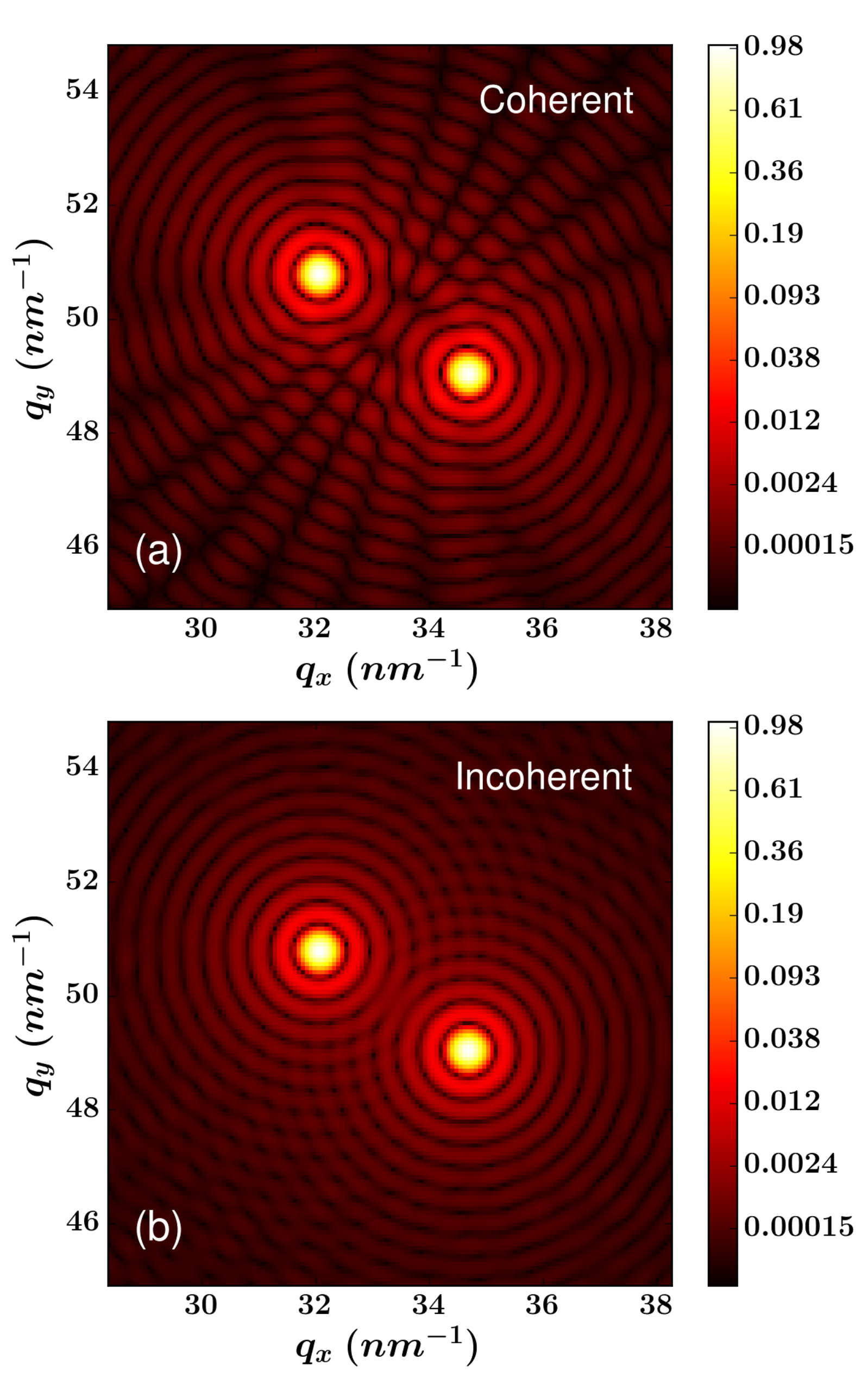}
\caption{(a) Simulated coherent X-ray diffraction from two Copper cylinders 
(height 16nm, diameter 16nm) with one rotated by 3 degrees around their common
[011] axis.  The reciprocal space maps are in the (011)* plane around the
$2\overline{2}\overline{2}$ reciprocal space point. (b) Same as (a) but the
intensities diffracted by the two cylinders are added as if the two objects
were separated by a distance much larger than the X-ray beam coherence
length (and therefore scatter incoherently).}
\label{coherent}
\end{figure}

To illustrate this point, let us for instance consider the model system
envisaged by LRPS made of two identical Pd cylinders with the upper one
rotated by different angles around the common [hh0] axis. According to these
authors {\it ``A WPPM analysis of the corresponding powder patterns shows that
for tilt angles $>$ 1.5 $^{\circ}$ coherence between the two half-cylinders is
completely lost, so that powder diffraction ``sees'' completely separate
(incoherently scattering) domains''}. Considering now the same sample studied
with X-ray coherent diffraction, this supposed ``loss of coherence'' is not
observed. Figure \ref{coherent} shows an example of large tilt angle (3
degrees) where clearly one can make the difference between the real
diffraction pattern from the two cylinders [Fig.\ref{coherent}(a)] and the one
that would correspond to incoherent diffraction [Fig.\ref{coherent}(b)].

Finally, let us mention that Fig.\ref{coherent}(a) is only a slice of a three
dimensional reciprocal space structure. From the measurement of this latter,
associated with measurements around other reciprocal space points, an
inversion method should provide the two cylinders structure including their
relative orientation. This method was used recently to determine the structure
of inversion domains in a Gallium Nitride nanowire\cite{Stephaneacsnano}, a
system which presents similarities with the one discussed here.

%


\end{document}